# Tunable Fano resonances in heterogenous Al-Ag nanorod dimer

Xueting Ci, Botao Wu, Min Song, Yan Liu, Gengxu Chen, E Wu and Heping Zeng

State Key Laboratory of Precision Spectroscopy, East China Normal University, Shanghai, 200062, China

E- mail: btwu@phy.ecnu.edu.cn and hpzeng@phy.ecnu.edu.cn

## Abstract

We theoretically investigate the plasmonic coupling in heterogenous Al-Ag nanorod dimer. A pronounced Fano dip is found in the extinction spectrum produced by the destructive interference between the bright dipole mode from a short Al nanorod and the dark quadrupole mode from a long Ag nanorod nearby. This Fano resonance can be widely tuned in both wavelength and amplitude by varying the rod dimensions, the separation distance and the local dielectric environment. The Al-Ag heterogeneous nanorod dimer shows a high sensitivity to the surrounding environment with a local surface plasmon resonance figure of merit of 7.0, which enables its promising applications in plasmonic sensing and detection.

# 1. Introduction

The coupling between different plasmon modes in plasmonic complex nanostructures such as closely placed metallic nanoparticles has been a flourishing research field with great importance in recent years [1]. When interacting with light, the metallic complex nanostructures stand out the ability to localize and manipulate the electromagnetic waves down to nanometer scale due to the collective oscillation of free conduction electrons on the metal surface, which is known as localized surface plasmon resonance (LSPR). The resonant excitation of LSPR in the metallic complex nanostructures can induce extremely strong localized near field intensity enhancement due to the near filed coupling between adjacent nanoparticles, and has enabled a rich variety of applications in plasmonic photovoltaic cells, surface-plasmon enhanced spectroscopies, photochemistry, sensors, photodetectors, and quantum optics [2-12]. Furthermore, LSPR properties of metallic complex nanostructures can be tuned by varying the metal composition, shape, size, and surrounding environment.

Among various novel plasmonic phenomena, Fano resonance in metallic complex nanostructures from the destructive interference between a broad super-radiant bright mode and a narrow sub-radiant dark mode has attracted extensive research attention due to its narrow and asymmetric lineshape and sensitivity to structural and environmental parameters, showing potential applications in plasmonic sensing and surface enhanced Raman scattering [13-15]. Fano resonance has been extensively studied in many nanoparticle aggregates formed by the same metal, in which the nanoparticle size, spacing between the nanoparticles, symmetry of the aggregates, and incident light polarization directions all give rise to the tuning of Fano resonance [13-21]. Fano resonance has also been reported to a lesser extent in heterogenous nanostructures constituted by nanoparticles of different metals, theoretically and experimentally, mainly concentrating on Au-Ag heterodimers [22-27]. In this paper, we present a theoretical investigation on the Fano resonance in a new kind of heterogeneous metallic nanostructure by Al-Ag nanorod dimers.

As shown in figure 1(a), the Al and Ag nanorods are arranged by end-to-end with a small gap in between. The Al nanorods have been recently reported to exhibit highly tunable plasmonic resonances from the deep ultraviolet to the visible wavelength region [28-30]. Here, the short Al nanorod supports a broad bright mode while the long Ag nanorod supports a narrow dark mode. The interaction between them induces an apparent Fano dip in the extinction spectra of Al-Ag nanorod dimers. The Fano resonance can be tuned by changing the Al or Ag nanorod length and spatial separation in between, and shows a high sensitivity to the surrounding environment.

## 2. Simulation method

The numerical simulations were performed by the finite difference time domain (FDTD) Solutions software (Version 8.5, Lumerical Solution, Inc. Canada). The dielectric constants of aluminum and silver are taken from [31]. The schematic geometry of Al-Ag nanorod dimer is shown in figure 1(a). The Al and Ag nanorods are coaxially arranged with the lengths of $L_1$ and $L_2$, respectively and the diameter $2R$. The gap between the two nanorods is $d$. Supposing that the Al-Ag nanorod dimer is placed in the air, the refractive index of the surrounding matrix is set to be 1. A plane wave total field-scattered field source ranging from 200 to 900 nm is utilized as the incident light beam with linear polarization direction along the longitudinal axis of the nanorod dimer shown as $x$ direction in figure 1(a). A three-dimensional nonuniform meshing is used, and a grid size of 0.5 nm is chosen for the inside and immediate vicinity of Al-Ag nanorod dimers. We use perfectly matched layer absorption boundary conditions as well as symmetric boundary conditions to reduce the memory requirement and computational time. All the numerical results pass prior convergence testing.

## 3. Results and discussion

Firstly we investigate how to produce the Fano resonance between the bright and dark modes supported by single Al and Ag nanorods by the destructive interference. Figure 1(b) presents the extinction spectrum of one single Al nanorod with the length $L_1 = 142$ nm and the diameter $2R = 40$ nm. The resonance peak around 515 nm, originating from a dipole surface plasmon resonance, can serve as a bright mode. Meanwhile, the dark mode cannot be excited by the plane wave due to their mutual weak coupling, but can be excited by a point source [19]. Figure 1(c) displays the extinction spectrum and nonradiative enhancement of one single Ag nanorod corresponding to the normal plane wave illumination and dipole source excitation, respectively. The diameter of Ag nanorod is also $2R = 40$ nm and the length is $L_2 = 200$ nm. The dipole source is placed along the longitudinal axis of the rod with the distance $d = 10$ nm between the dipole source and the end face of the rod, and the polarization of the dipole source is also along the longitudinal axis of the rod [the inset in figure 1(c)]. It can be found that there is a resonance peak around 512 nm induced by the excitation of the dipole source. This cannot be excited by the normal plane wave, since it is a quadrupole plasmon resonance and can serve as a dark mode. Obviously, if the dipole source is replaced by the Al nanorod the dark mode in the Ag nanorod will be excited under the illumination of the plane wave, and concomitantly Fano resonance will appear due to their matched resonance frequency.

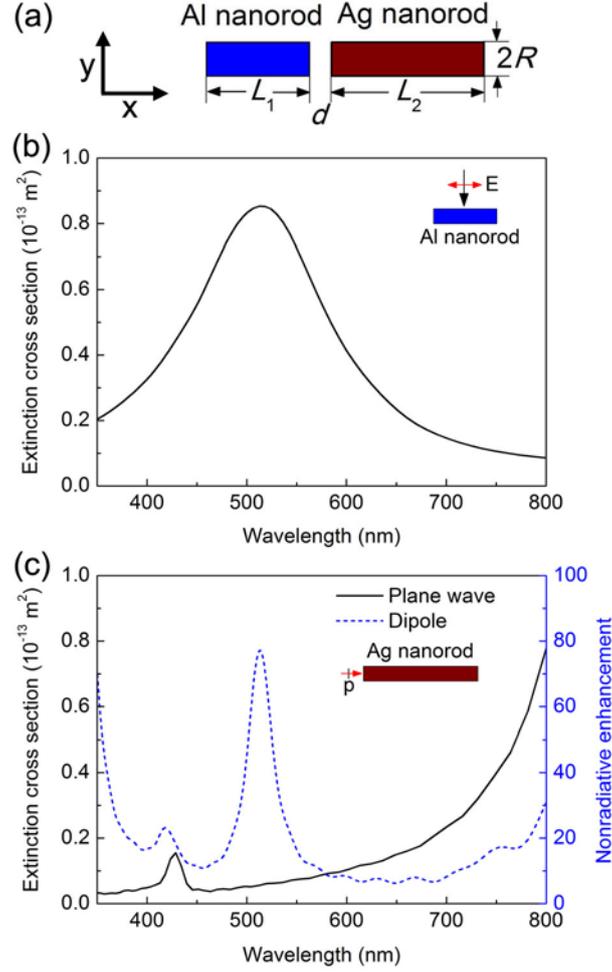

**Figure 1.** (a) Geometry of an Al-Ag heterogenous nanorod dimer. (b) Extinction spectrum of a short Al nanorod with $L_1 = 142$ nm and $R = 20$ nm. The inset shows the schematic of the nanorod illuminated by a plane wave. (c) Extinction spectrum (black curve) and nonradiative enhancement (blue curve) of a long Ag nanorod with $L_2 = 200$ nm and $R = 20$ nm. The inset shows the configuration of the dipole-nanorod coupling system.

Figure 2(a) shows the extinction, scattering and absorption spectra of Al-Ag heterogenous nanorod dimer consisting of the above mentioned Al and Ag nanorods with $d = 10$ nm. Clearly, two resonant peaks ($\lambda = 473$ and 567 nm) with a dip at 512 nm appear in the extinction spectrum. Actually the dip is caused by Fano interference. According to [18], if one system is pumped at frequencies resonant with both bright and dark modes, the bright mode will be excited by two pathways: $|I\rangle \to |B\rangle$ and $|I\rangle \to |B\rangle \to |D\rangle \to |B\rangle$, where $|I\rangle$, $|B\rangle$ and $|D\rangle$ are excitation source, bright mode, and dark mode, respectively. Fano-like interference occurs when the cumulative phase shift from $|B\rangle \to |D\rangle \to |B\rangle$ is π so that the two pathways interfere destructively, canceling the polarization of the bright mode and resulting in a narrow Fano dip in the extinction spectrum as shown in figure 2(a). Simultaneously, the dipole and quadrupole plasmon modes supported by the short Al and long Ag nanorods couple in a constructive way to form two collective resonant modes: a high-energy antibonding mode at 473 nm and a low-energy bonding mode at 567 nm. The electric field distributions of the dimer with the separation $d$ = 10 nm at the wavelength of 473, 512, and 567 nm are shown in figures 2(b)-2(d). The electric field enhancement of the dimer at the Fano dip is greatly depressed due to the cancelation of its polarization in the Fano resonance [see figure 2(c)].

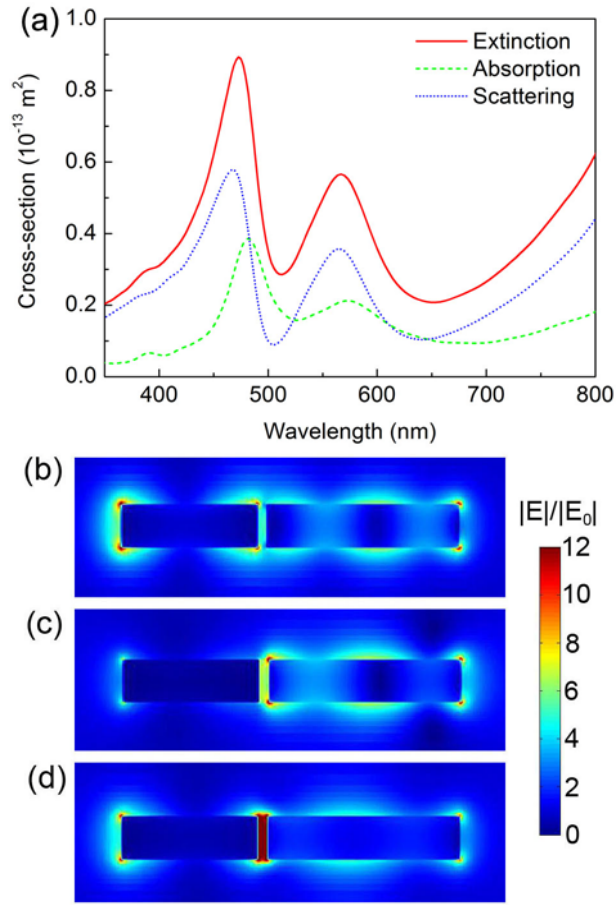

**Figure 2.** (a) Extinction, absorption and scattering spectra of an Al-Ag nanorod dimer with $L_1$ = 142 nm, $L_2$ = 200 nm, and $R$ = 20 nm. Electric field distributions across the central cross section of the Al-Ag nanorod dimer at $\lambda$ = 473 nm (b), 512 nm (c), and (d) 567 nm.

The Fano interference strength in Al-Ag nanorod dimer can be changed by modulating the gap between the two nanorods as shown in figure 3. Decreasing the separation between the Al and Ag nanorods enhances the coupling between the bright and dark modes, and concomitantly the Fano dip becomes deeper [19]. The deeper Fano dip may be useful in the application of plasmon induced transparency [32]. In addition, the variation of plasmon coupling strength shows more apparent effect on the low-energy peak ($\lambda$ = 567 nm) than that on the high-energy one ($\lambda$ = 473 nm). With the gap $d$ increasing from 10 to 30 nm, the low-energy peak shows a blue shift of about 25 nm (from 567 to 542 nm), whereas the high-energy peak shows a red shift of about 11 nm (from 473 to 484 nm). In the case of Fano dip, it only shifts from 512 to 510 nm.

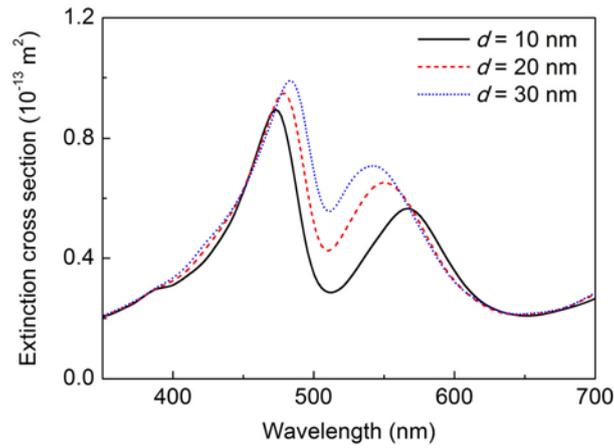

**Figure 3.** Extinction spectra of Al-Ag nanorod dimers with gap distances $d$ = 10, 20 and 30 nm, respectively. The lengths of Al and Ag nanorods are 142 and 200 nm, respectively, and the radius of each rod is 20 nm.

The Fano resonance character can also be tuned by adjusting the rod length of the short Al nanorod (bright mode) or the long Ag nanorod (dark mode), as shown in figure 4. In figure 4(a), as the length of the short Al nanorod increases from 132 to 192 nm with fixed length of Ag nanorod of 200 nm, the long wavelength resonant peak intensity relative to the Fano dip significantly increases while the short wavelength one gently decreases. In addition, the bandwidth of the long wavelength resonant peak largely broadens while the short wavelength one slightly narrows. When increasing the length of the long Ag nanorod from 150 to 230 nm with fixing the short Al nanorod length at 142 nm, the extinction peak intensity and the bandwidth of both two peaks exhibit an opposite behaviors comparing with the case of the length increase of the short Al nanorod [see figure 4(b)].

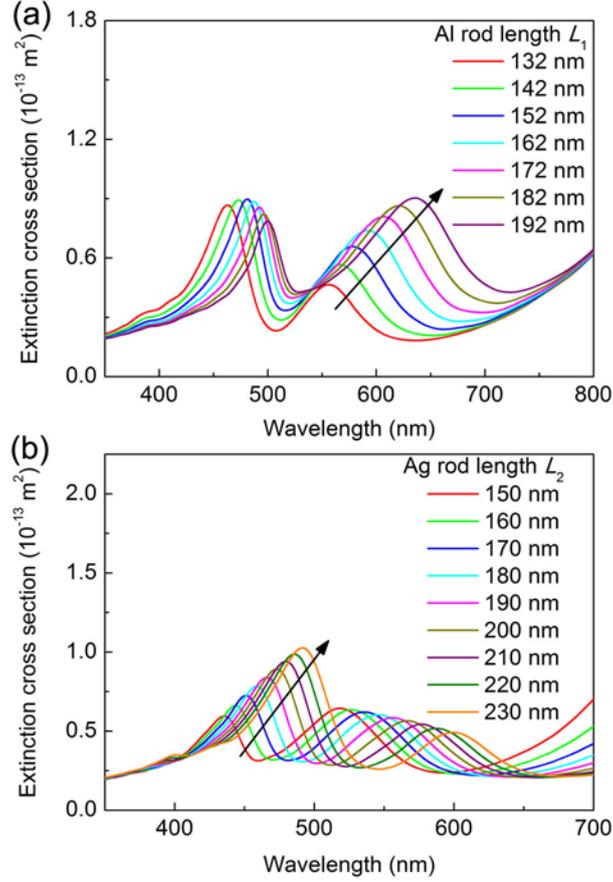

**Figure 4.** (a) Extinction spectra of Al-Ag nanorod dimers with different Al nanorod length $L_1$, and other parameters are fixed at $L_2$ = 200 nm and $d$ = 10 nm. (b) Extinction spectra of Al-Ag nanorod dimers with different Ag nanorod length $L_2$, and other parameters are fixed at $L_1$ = 142 nm and $d$ = 10 nm. The arrows in (a) and (b) indicate the incremental directions of the Al and Ag rod length, respectively.

One of the very interesting potential applications for Fano resonance in plasmonic systems is biological sensors, which stems from its LSPR inherent sensitivity to the change of the local surrounding environment. The efficiency of LSPR sensors is typically evaluated by its figure of merit (FoM), defined as:

$$FoM = \frac{m(eV/RIU)}{FWHM} \quad (1)$$

where $m$ is the plasmon energy change per refractive index unit, and FWHM is the full width at half maximum of the resonant peak [33]. In order to obtain a high FoM for the Al-Ag nanorod dimer, it is necessary to consider the balance of the extinction cross section and FWHM of a resonance. Larger extinction cross section will result in a better sensing performance, but FWHM will be broadened at the same time, which will reduce the FoM. Here, we choose the Al-Ag nanorod dimer with the same size parameters as those in figure 2(a) to simulate its LSPR sensitivity to the refractive index change of the surrounding medium, because its antibonding mode at 473 nm shows a large extinction cross section and a narrow FWHM. As shown in figure 5(a), with the environment refractive index changing from 1.0 to 1.5, the peak of the antibonding mode moves from 473 to 672 nm. A slope of about 1.566 is obtained by a linear fit for the peak energy shift of the antibonding mode as a function of the refractive index of the surrounding environment [figure 5(b)]. Using the above mentioned equation, the slope is divided by the peak FWHM 81 nm (0.223 eV) and the resulting FoM of about 7.0 is obtained, which is larger than those of isolated nanoparticles such as nanocubes (FoM = 5.4) [33] and nanoclusters (FoM = 5.7) [17].

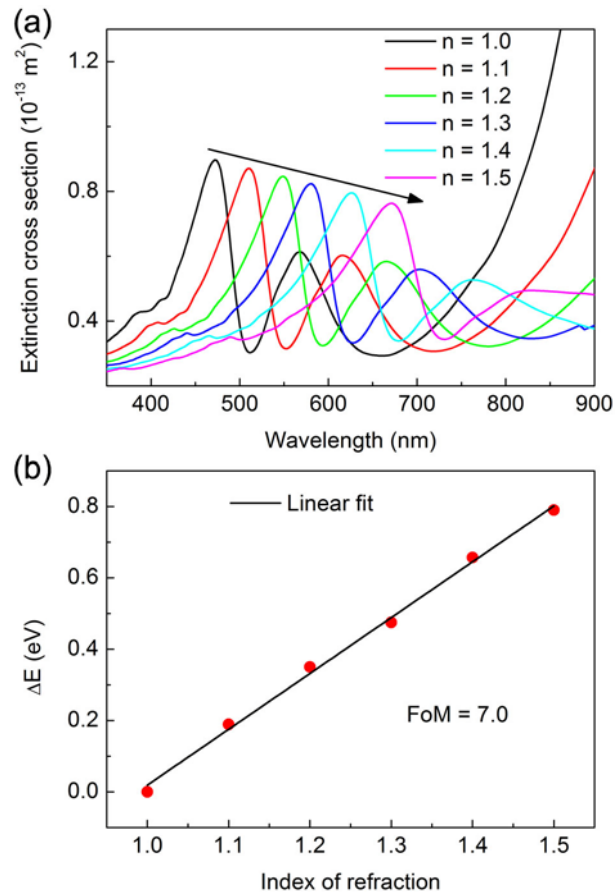

**Figure 5.** (a) Extinction spectra of Al-Ag nanorod dimer with the environment refractive index *n* of 1.0, 1.3, 1.4 and 1.5, respectively. The arrow indicates the incremental directions of the refractive index. (b) Linear plot of LSPR peak shifts at 473 nm vs refractive index of the surrounding environment.

## 4. Conclusions

In conclusion, optical properties of Al-Ag heterogeneous nanorod dimers are studied by theoretical simulation method. A pronounced Fano dip in the extinction spectra is observed, which strongly depends on both the geometry parameters of the complex nanostructure and the refractive index of the surrounding environment. The LSPR sensitivity of the heterogeneous metallic complex nanostructures is also checked and a FoM of 7.0 is obtained, which may find applications in biological sensing and molecule detection based on the coherent plasmonic coupling.

## Acknowledgements

This work was funded in part by National Nature Science Fund (11104079), Research Fund for the Doctoral Program of Higher Education of China (20110076120019), National Key Scientific Instrument Project (2012YQ150092), and the State Key Laboratory of Luminescent Materials and Devices at South China University of Technology.